\documentclass[prd,amsfonts,twocolumn,nofootinbib,showpacs]{revtex4}
\RequirePackage{graphicx}
\usepackage{enumerate}
\usepackage{comment}

\begin{document}
\title{Preheating with higher dimensional interaction}

\author{Seishi Enomoto}
\affiliation{Institute of Theoretical Physics, Faculty of Physics,
University of Warsaw, ul.Pasteura 5, 02-093 Warsaw, Poland}
\author{Nobuhiro Maekawa}
\affiliation{Kobayashi Maskawa Institute, Nagoya University; \\
Department of Physics, Nagoya University, Nagoya 464-8602, Japan}
\author{Tomohiro Matsuda}
\affiliation{Laboratory of Physics, Saitama Institute of Technology, Fukaya, Saitama 369-0293, Japan}

\begin{abstract}
Particle production caused by the oscillation after inflation is important
 since it explains reheating after inflation.
On the particle theory side, we know that effective action may have
 additional higher dimensional terms (usually 
 called non-renormalizable terms) suppressed by the cut-off scale.
Moreover, interaction between inflaton and so-called moduli field
 will be higher dimensional.
Therefore, if such higher dimensional interaction is significant for resonant
 particle production, one cannot avoid the effect in preheating study. 
We explicitly calculated the required number of oscillation
for the energy transfer.
Consequently, cosmological history of an oscillating field and the
 moduli problem can be reconsidered.
\end{abstract}

\maketitle

\section{Introduction}
If the Universe starts with inflation, elementary particles of the
Universe should be created after inflation.
According to the inflationary theory, reheating after inflation, 
which converts the vacuum energy into radiation, is
responsible for the creation of those particles.
At the beginning of that process, inflaton field is supposed to start
oscillation, and then reheating occurs due to particle production caused
by the oscillating field.
The oscillation causes production of particles, which interact with each
other and finally they come to be in thermal equilibrium.
The mechanism of particle production after inflation has been discussed
by many authors~\cite{Review-paper}. 
In this paper we consider a case in which reheating starts with a regime
called preheating~\cite{TowardPR}.
In Ref.\cite{TowardPR}, assuming a renormalizable interaction, it has
been shown that during that period the energy transfer from the inflaton
oscillation to other Bose fields and particles is extremely efficient.
In this paper we are focusing on the initial stage, where preheating
occurs and particles are produced.
Cosmological evolution of the Universe after preheating could be
different from the one after perturbative reheating.

In contrast to the generality of the mechanism, the original scenario of
preheating is based on renormalizable interaction, which restricts the
application of the idea only to the fields that have renormalizable 
interaction.
Therefore, although many theoretical models of particle physics may have
fields that do not have renormalizable interaction with inflaton, 
preheating has been applied only to those ``special'' cases.
Although not mandatory, inflationary model may expect amplitude as large
as the Planck scale~\cite{BICEP2, Lyth-bound} or the scale of Grand
Unified Theory (GUT).

Before discussing the calculational details, it will be helpful to set out 
naive questions that may arise when one considers the case. 
An important point is that there has been no paper discussing
higher dimensional interaction for preheating.
Obviously such interaction has been neglected, but the reason is not
quite obvious. 
\begin{enumerate}
\item One might claim that large amplitude cancels the suppression of the
      cutoff. On the other hand, one might wonder whether such speculation
      can be applied when particle production occurs near the origin. 

\item One might think that preheating could be negligible when the
      coupling is very weak. Since higher 
      dimensional interaction is suppressed by the cutoff, it is usually
      ``very weak'' compared with renormalizable interaction. Then it
      could be natural to conclude that higher dimensional interaction is
      negligible.
\item In contrast to 1.$\sim 2,$ one might have an intuition that
      particle production will be trivial and there will be no
      significant difference from the 
      standard scenario.
      Then, one might simply assume that only the critical
      frequency where the infrared instability band ends will change.
\end{enumerate}
Also, if one focuses on particle production at the enhanced
symmetric point (ESP), one might have an impression that preheating
might not work since higher dimensional interaction gives
$\dot{m}_\chi\simeq0$, where $\chi$ is the scalar field that is supposed to
be produced.
This could not be a problem since the adiabatic condition is broken
when the inflaton moves a small distance away from the ESP, but
obviously the situation is not trivial.

In this paper we carefully consider these naive
speculations and intuitions.
We also evaluate numerical coefficients since these coefficients are
very important for quantitative estimation of the number density.
They appear in recurrence formula of geometric series.
Quantitative calculation of resonant particle production in
the expanding Universe is not obvious.
This could be in contrast to the transparency of the argument based on
the Mathieu equation or Floquet theory~\cite{Math-text}. 

Recently, we considered higher dimensional interaction for a trapping
mechanism~\cite{EMM} and obtained a result: higher dimensional
interaction can cause resonant particle production, and the particles can
induce significant trapping force.
Then, a natural question arises.
``Is the calculation applies to preheating after inflation? Is the
higher dimensional interaction causes resonant particle production when
a curvaton or a moduli is oscillating?''
This paper aims to gives an answer to those questions.

\section{What are the differences?}
In this section we carefully review the standard scenario of
preheating~\cite{TowardPR, InstantPR} to show which equations have to be 
modified.
The starting point of the ``standard'' preheating scenario is the Lagrangian
\begin{equation}
{\cal L}=\frac{1}{2}\partial_\mu\phi\partial^{\mu}\phi
+\frac{1}{2}\partial_\mu\phi\partial^{\mu}\chi
 -\frac{1}{2}m^2 \phi^2
-\frac{1}{2}g^2\phi^2\chi^2.
\end{equation}
Assuming homogeneous background, classical equation 
of $\phi$ obeys
\begin{equation}
\ddot{\phi}+3H\dot{\phi}+\left(m^2+g^2\left<\chi^2\right>\right)\phi=0.
\end{equation}
Here $\left<\chi \right>$ is initially negligible if $\phi$-oscillation starts with
large amplitude $\Phi_0$.
Then, interaction between the classical field $\phi$ and the
quantum scalar field $\chi$ is considered.
One can write the equation for modes with physical momentum $k$ in the
form
\begin{equation}
\label{eq-basic1}
\ddot{\chi}_k+(k^2+g^2\Phi^2\sin^2(mt))\chi_k=0
\end{equation}
This equation describes an oscillator with a frequency
$\omega^2=k^2+g^2\Phi^2 \sin^2(mt)$.
If $\Phi$ is {\bf not time-dependent} (i.e, when $\Phi=\Phi_0$), the equation
can be written as a Mathieu equation with $A_k\equiv \frac{k^2}{m^2} +2q$,
$q\equiv \frac{g^2\Phi^2}{4m^2}$, $z\equiv mt$.
Mathieu equation gives so-called ``broad resonance'' when $q\gg1$.
For very small $\phi$ the change in $\omega$ ceases to be adiabatic.
The standard condition necessary for particle production is
\begin{equation}
\frac{\dot{\omega}}{\omega^2}> 1.
\end{equation}
Therefore, the typical momenta for the resonance is
estimated as $g^2\phi\dot{\phi}<\omega^3$ for $\dot{\phi}\simeq m\Phi$, or
\begin{equation}
k^2 <k^2_*\equiv  (g^2\phi m\Phi )^{2/3}-g^2\phi^2.
\end{equation}
The maximum $\phi$ is (which appears when $k=0$)
\begin{equation}
\phi<\phi_*\equiv \frac{1}{2}\sqrt{\frac{m\Phi}{g}}.
\end{equation}
Thinking about higher dimensional interaction, which (in the simplest 
case) can be written as $\sim g^2 \phi^4\chi^2/\Lambda^2$,
it is easy to find that Eq.(\ref{eq-basic1}) will be replaced by
\begin{equation}
\label{eq-basic1high}
\ddot{\chi}_k+\left(k^2+g^2\frac{\Phi^4}{\Lambda^2}\sin^4(mt)\right)\chi_k=0.
\end{equation}
Although $\sin^2(mt)$ has been replaced by $\sin^4(mt)$, Floquet theory
predicts (rather naively) resonance and exponential growth.
Therefore, intuitive argument suggests that broad resonance will happen
for higher dimensional interaction if the Universe is not expanding.
On the other hand, quantitative calculation of the resonant particle
production in an expanding Universe is not obvious.
Our paper aims to find both analytical and numerical estimation of the
effect in the expanding Universe.

\section{Model and calculation}
In this section we carefully follow the calculation in
Ref.\cite{TowardPR, CurvatonPR} so that the reader can easily compare
our result with the usual calculation.
Our formulation is carefully prepared so that one can easily
figure out why preheating is efficient even though the interaction is
higher dimensional.
For that purpose we start with an obvious case using a rather
special approximation, which is called ``quadratic approximation'' in
this paper.

Numerical coefficients are carefully evaluated so that it does not ruin the
accuracy of the estimation.
More detailed argument on application to the effective action of GUT
will be found in Ref.~\cite{EMM}. 
Note that our calculation has direct application to those
phenomenological models.

\subsection{Rather trivial example: quadratic approximation}
Our aim in this section is to find an obvious example of higher
dimensional interaction that can be examined using the standard method
of preheating.
Although the approximation is valid only in a very narrow range of the
parameter space, the result gives a convincing lower limit of the
particle production.
We are going to extend the analysis in the next section, using more
rigorous calculation of Ref.\cite{EMM}.

First, consider a model of parametric resonance described by an oscillating
field $\phi$ and a particle $\chi$ with the potential and the
interaction:
\begin{equation}
V(\phi,\chi)=\frac{1}{2}m_\phi^2|\phi|^2+
\frac{g^2}{2} \frac{|\phi|^4}{\Lambda^2}\chi^2.
\end{equation}
Around j-th zero crossing at $t=t_j$, we assume sinusoidal oscillation
$\phi(t)=\Phi_j\sin[m_\phi (t-t_j)]+i\mu_j$, which leads to
\begin{eqnarray}
|\phi(t)|^2&=&\Phi^2_j\sin^2[m_\phi (t-t_j)]+\mu_j^2,
\end{eqnarray}
where $\Phi_j$ is the amplitude of the oscillation and $\mu_j$
is the impact parameter ($\mu_j \ll \Phi_j$).
Then one can linearize the interaction
using $\sin[m_\phi (t-t_j)]\simeq m_\phi(t-t_j)\ll 1$, and disregard
higher terms.
More explicitly, one will find
\begin{eqnarray}
|\phi|^4&=&\left[\Phi_j^2\left\{m_\phi (t-t_j)-\frac{1}{3!}m_\phi^3 (t-t_j)^3+...
\right\}^2+\mu_j^2\right]^2\nonumber\\
&=&\mu_j^4+ 2\mu_j^2 \Phi_j^2m_\phi^2 (t-t_j)^2+\Phi_j^4m_\phi^4(t-t_j)^4+...,
\end{eqnarray}
where quartic term ($\propto (t-t_j)^4$) can be neglected when 
$m_\phi(t-t_j)< \sqrt{2}\mu_j /\Phi_j$.
Since the velocity at the bottom of the potential is
$\dot{\phi}_j\simeq m_\phi\Phi_j$, the condition $m_\phi(t-t_j)< \sqrt{2}\mu_j
/\Phi_j$ is equivalent to  $|\mathrm{Re}\, \phi| < \sqrt{2}\mu_j$. 
For our estimation we will take $\mu_j$ modestly large so that 
the quadratic approximation will be conceivable.
Since a large $\mu_j$ will give an exponential suppression of the number
density, $\mu_j$ has to be chosen carefully.
Our choice of $\mu_j$ will be discussed in the last part of this section.
Explicit form of the higher dimensional interaction term is
\begin{eqnarray}
\label{quadra-approximation}
\frac{g^2}{2}\frac{|\phi(t)|^4}{\Lambda^2}\chi^2
&\simeq&
\frac{g^2}{2}\frac{2\mu_j^2\Phi(t)^2m_\phi^2 (t-t_j)^2+\mu_j^4}{\Lambda^2}\chi^2.
\end{eqnarray}
Remember that for a renormalizable coupling ($g^2 |\phi|^2\chi^2/2$), one
will find 
\begin{equation}
\frac{g^2}{2}|\phi(t)|^2 \chi^2 \simeq
\frac{g^2}{2}\left\{\Phi(t)^2 m_\phi^2 (t-t_j)^2+\mu_j^2\right\} \chi^2,
\end{equation}
which does not have the suppression ($\sim\frac{\mu_j^2}{\Lambda^2}$) in
front of the effective mass term.

For a flat Friedmann background with
cosmological scale factor $a(t)$, the equation of motion of the field
 $\chi(\mathbf{x},t)$ in Fourier space of comoving momentum $\mathbf{k}$ is
written for  $\chi_k(t)$ as
\begin{eqnarray}
&&\ddot{\chi}_k+3H\dot{\chi}_k\nonumber\\
&+&\left(
\frac{2g^2 \mu_j^2\Phi_j^2m^2_\phi (t-t_j)^2+g^2\mu_j^4}{\Lambda^2}
+\frac{\mathbf{k}^2}{a^2(t)}
\right)\chi_k\nonumber\\
&=&0.
\end{eqnarray}
The physical momentum $\mathbf{p}=\mathbf{k}/a(t)$ coincides with
$\mathbf{k}$ for Minkowski space.
We can eliminate the friction term $3H\dot{\chi}_k$ by defining 
$X_k\equiv a^{3/2}\chi_k$.
Then, one can rewrite the equation in a simpler form
\begin{eqnarray}
\label{eq:eom_chi}
\ddot{X}_k
+\left(P^4_j(t-t_j)^2+A_j
+\frac{\mathbf{k}^2}{a^2(t)}\right)X_k&=&0,
\end{eqnarray}
where $P_j^2\equiv \sqrt{2}g\mu_j\Phi_j m_\phi/\Lambda$.
Changing the time variable as $\tau\equiv P_j (t-t_j)$, the equation gives
\begin{equation}
\label{basic-eq}
X_k''+ (\kappa^2_j+\tau^2)X_k=0,
\end{equation}
where prime denotes derivatives with respect to $\tau$.
$\kappa_j$ is defined by
\begin{equation}
\kappa_j^2\equiv \left(\frac{k}{k_{*,j}}\right)^2
+\frac{a(t_j)^2}{k_{*,j}^2} A_j,
\end{equation}
where $k_{*,j}\equiv a(t_j) P_j$.
The equation of motion contains terms
proportional to $H^2X_k$ and $\frac{\ddot{a}}{a}X_k$,
which are included in $A_j$.
One will see that these terms can be neglected after
all, since $k^2/a^2\gg H^2, \ddot{a}/a$ at subhorizon scales~\cite{TowardPR,CurvatonPR}. 
The remaining term will be $A_j\simeq g^2\mu_j^4/\Lambda^2$.

Eq.(\ref{basic-eq}) can be solved as the well-known problem of wave scattering
at a negative parabolic potential, which leads to 
\begin{eqnarray}
\label{result1}
n_{k,j}&=&e^{-\pi\kappa_j^2}+(1+2e^{-\pi\kappa_j^2})n_{k,j-1}\nonumber\\
&&-2e^{-\frac{\pi}{2}\kappa_j^2}\sqrt{1+e^{-\pi\kappa_j^2}}
\sqrt{n_{k,j-1}(1+n_{k,j-1})}\sin\theta_j,\nonumber\\
\end{eqnarray}
where the phase $\theta_j$ causes stochastic growth of the occupation
number $n_k$.
According to Ref.\cite{TowardPR}, the stochastic contribution can be
averaged to zero after all, since $\theta_j$ is uniformly distributed.
Then one will find the iterative expression for the occupation number
$n_k$ as~\cite{CurvatonPR}
\begin{eqnarray}
\label{result1-n}
\left(n_{k,j}+\frac{1}{2}\right)
&=&(1+2e^{-\pi\kappa_j^2})\left(n_{k,j-1}+\frac{1}{2}\right).
\end{eqnarray}
Since the occupation number becomes large soon after $\phi$ starts
oscillation, 
the equation shows that the mode occupation number $n_k$ increases
exponentially as long as the mode satisfies $\pi \kappa_j^2<1$.

Although the formalism might look completely the same as the conventional
calculation, a significant difference appears in the definition of
$k_*$, which defines the threshold of the particle production.
If the interaction is renormalizable, one will find 
$k_{*,j}^2/a(t_j)^2\sim gm_\phi\Phi_j\sim g\dot{\phi}$, while for higher dimensional
interaction one will find $k_{*,j}^2/a(t_j)^2\sim \sqrt{2}g\mu_j
m_\phi\Phi_j/\Lambda$.
In contrast to the case with $\mu\simeq 0$, $\dot{\omega}$ does not
vanish at the ESP.

\begin{figure}[t]
\centering
\includegraphics[width=1.0\columnwidth]{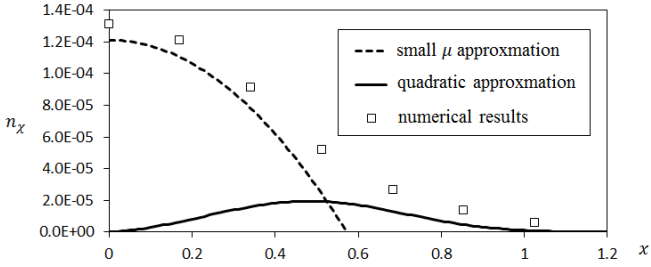}
 \caption{The picture shows our analytical and numerical
 calculations in Minkowski space for the first particle production ($n_{\chi,1}$).
 We took $g=1$, $m_\phi=0.1\Lambda$, and $\Phi_1=1.0\Lambda$.
 (Results are calculated for a flat potential ($m_{\phi}=0$)
 with initial condition $\dot{\phi}=0.1\Lambda^2$.)}
\label{fig:comparing}
\end{figure}

Let us examine the validity of the above calculation. 
The condition that is needed for the quadratic approximation is
$|{\rm Re} \phi|<\sqrt{2}\mu_j$.
At the same time, we need to consider the adiabatic condition
($\dot{\omega_k}/\omega_k^2<1$,
where $\omega_k = \sqrt{\mathbf{k}^2 + g^2|\phi|^4/|\Lambda^2}$),
which is violated when non-adiabatic particle production is efficient.
One can rewrite the adiabatic condition as 
\begin{eqnarray}
\label{adi-th}
|\phi|&>&\phi_{*j}\equiv\left(\frac{2\Lambda \dot{\phi}_j}{g}\right)^{1/3}.
\end{eqnarray}
Requiring $|{\rm Re} \phi|<\sqrt{2}\mu_j$ whenever
the adiabatic condition is violated, 
a lower bound for $\mu_j$ will be obtained, which  is given by a
parameter
\begin{equation}
x_j \equiv \mu_j/\phi_{*j}>1/\sqrt{3} \sim 0.577.
\end{equation}
Here we will focus on the first particle production and discuss the
validity of the approximation.
Then in the next section we will discuss the resonant particle production.
In our formalism, the produced number density can be estimated as
\begin{eqnarray}
n_{\chi,1} &=& \frac{1}{2\pi^2a(t_1)^3}\int_0^{\infty} dk k^2 n_{k,1} \nonumber \\
 &=& \frac{2^{5/4}}{(2\pi)^3} \frac{g \Phi_1^2 m_{\phi}^2}{\Lambda} x_1^{3/2} e^{-\sqrt{2}\pi x_1^3}.
\end{eqnarray}
In Ref.~\cite{EMM}, we have another method of analytical calculation, which is
valid for small $\mu$. 
It shows
\begin{eqnarray}
\label{EMM-eq}
 n_{\chi} &\sim&  0.0121 \cdot \frac{g\dot{\phi}^2}{\Lambda}
  \left( 1 - 3.06 x_1^2 + \mathcal{O}(x_1^4) \right),
\end{eqnarray}
where one can substitute $\dot{\phi} \sim \Phi_1 m_{\phi}$ if $\Phi_1 \gg m_{\phi}$.
Our analytical and numerical results are compared in Fig.\ref{fig:comparing},
in which one will see clearly that the quadratic approximation is
conceivable when $x_1>0.577$, 
while analytical calculation in Ref.~\cite{EMM} (small $\mu$
approximation) is conceivable when
$x_1<1/\sqrt{3.06}=0.572$.
(Higher terms in Eq.(\ref{EMM-eq}) start to dominate at $x_1=0.572$.)

\subsection{Growth of number density and energy transfer in quadratic approximation}
In this section we mainly follow the useful calculational method
considered in Ref.\cite{CurvatonPR} and estimate the number of
oscillations needed for significant energy transfer.

When the Universe is dominated by matter, or by oscillating field that
scales like matter, $\Phi_j$ and $a_j$ depend on $j$ as~\cite{CurvatonPR}\footnote{Alternatively, one may consider quartic inflaton potential $V(\phi)\sim g^2 \phi^4$ that
leads to non-sinusoidal oscillation and radiation-like evolution of the energy
density, which (approximately) gives $\Phi_j \propto j^{-1}, a_j\propto
j$~\cite{quartic-pr}. }
\begin{equation}
\Phi_j \propto j^{-1},\qquad
a_j\propto j^{2/3}.
\end{equation}

If we take a fixed value $x_j=x$ (for example, $x=0.577$)  for all $j$, 
we find
\begin{equation}
\label{eq:k_j_dependence}
k_{*,j} \propto a_j \mu_j^{1/2} \Phi_j^{1/2} \propto
a_j \phi_{*,j}^{1/2} \Phi_j^{1/2}\propto j^0,
\end{equation}
which leads to $e^{-\pi \kappa_j^2}\sim e^{-\pi\kappa_1^2}=n_{k,1}$.
Then, Eq.(\ref{result1-n}) gives
\begin{eqnarray}
n_{k,j}
&\simeq&  (1+2 n_{k,1})n_{k,j-1}+n_{k,1}\nonumber\\
&\simeq&\frac{1}{2}\sum_{I=1}^j {}_j{\mathrm C}_I (2n_{k,1})^I, \label{eq:nkj}
\end{eqnarray}
where ${}_j{\mathrm C}_I$ is binomial coefficient.

One might suspect that the fixed value $x_j\sim 0.6$ could be incorrect in reality, since 
$x_j\sim0.6$ cannot be applied for every $j$ at the same time.
In fact, $x_j$ will be proportional to $j^{-2/3}$ because $\mu_j \propto j^{-1}$
(when the quantity $a(t_j)^3 \Phi_j \mu_j$ is conserved),
while $\phi_{*j}$ will be proportional to $\Phi_j^{1/3} \propto j^{-1/3}$.
This means that $x_j\sim 0.6$ will be broken soon after the
first particle production. 
If one wants to see the particle production in the inner area
($x_j<x=0.577$),  Fig.\ref{fig:comparing} will be a useful guide, in which
one will see that smaller $x_j$ enhances particle production. 
Therefore, although the estimation of Eq.(\ref{eq:nkj}) is not accurate
when $x\sim 0$, one can expect that the number density calculated for
$x_j\sim 0.6$ will give a conceivable lower bound for the particle production.

Assuming that $\chi$ does not decay until $t=t_j$,
the total energy transfer to $\rho_\chi$ is calculated as
\begin{eqnarray}
\label{eq-first-rho}
\rho_{\chi,j}&=& \frac{1}{2\pi^2 a(t_j)^3}\int dk k^2 \omega_k(t_j)
 n_{k,j}\nonumber\\
&\simeq&\frac{g\Phi_j^2}{\Lambda}\sum_{I=1}^j\frac{{}_j{\mathrm C}_IR^I}
{4\pi^2 a(t_j)^3}\int^{\infty}_0 dk k^2e^{-\pi
I\frac{k^2}{k_{*,1}^2}}\nonumber\\ 
&\simeq&\frac{g\Phi_1^2}{j^2\Lambda}\frac{2^{1/4}x_1^{3/2}}{8\pi^3}
\frac{g\dot{\phi}_1^2}{j^2\Lambda}\sum_{I=1}^j\frac{{}_j{\mathrm C}_IR^I}{I^{3/2}}\nonumber\\
&\simeq&\frac{2^{1/4}x_1^{3/2}}{8\pi^3}\frac{g^2\Phi_1^2}{j^2\Lambda^2}
\rho_{\phi,j}\sum_{I=1}^j\frac{{}_j{\mathrm C}_IR^I}{I^{3/2}}
\end{eqnarray}
where $R\equiv 2\exp\left({-\pi A_1 a(t_1)^2/k_{*,1}^2}\right)=2\exp 
\left(-\sqrt{2}\pi x_1^3\right)$.
If one needs to calculate the summation one can use
$n!\sim\sqrt{2\pi n}(n/e)^n$ together with saddle
point method for the integration $\sum_{I=1}^j\rightarrow\int_1^{j+1}dI$. 
For $x_1=0.577$ and $g\Phi_1\simeq \Lambda$,
$\rho_{\chi,j}\sim\rho_{\phi,j}$ will be achieved when  
$j\sim 25$.
Note however that $j\sim 25$ is a modest estimation.

\subsection{Apart from quadratic approximation}
\begin{figure}[t]
\centering
\includegraphics[width=1.0\columnwidth]{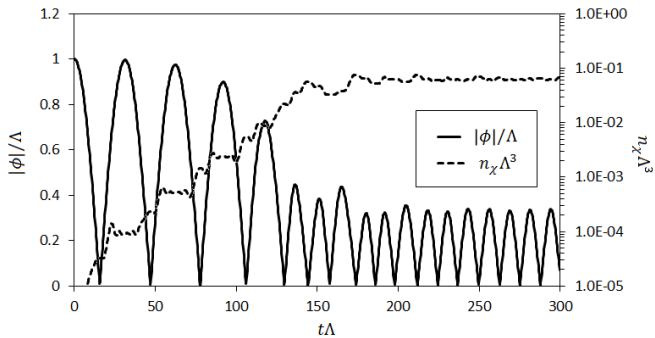}
 \caption{The picture shows our numerical calculation in Minkowski
 space with vanishing impact parameter ($\mu= 0$).
We took $\phi(t=0)=1.0\Lambda$, $\dot \phi(t=0)=0$, $g=1$, and $m_\phi=0.1\Lambda$.
 One will see that resonant particle production is efficient even if the
 quadratic approximation is not valid in the non-adiabatic region.}  
\label{enomoto-fig}
\end{figure}

In reality, one must take $x\ll 1$ and (intuitively) the energy transfer
will be more significant.
To confirm our intuition, we show our numerical calculation in
Fig.\ref{enomoto-fig}, which shows that $j=5$ is sufficient for the
energy transfer.
For simplicity we showed our numerical result calculated in
Minkowski space.
To compare it with our analytical calculation, we have to
 calculate the same quantity when the expansion of the Universe is
 neglected. 
In that case we can easily obtain $j\sim 13$ from the
analytical calculation (with quadratic approximation).
Moreover, already for the first particle production (see
Fig.\ref{fig:comparing}), the numerical calculation for $x=0$ shows more
significant particle production compared with the quadratic
approximation.

To improve our estimation, we propose another approximation that will
be better when $x_j\simeq 0$. 
Remember that Eq.(\ref{result1}), which explains resonant particle production,
is derived from the relation between Bogoliubov coefficients of the wave
function of $\chi$.
The original relation can be shown as~\cite{TowardPR}
\begin{equation}
\beta_{k,j} \sim e^{-\pi \kappa_j^2/2} \alpha_{k,j-1} + \sqrt{1+e^{-\pi \kappa_j^2}}
 \beta_{k,j-1}.
\end{equation}
The factors $e^{-\pi \kappa_j^2/2}$ and $\sqrt{1+e^{-\pi \kappa_j^2}}$
relate to the reflection and the transmission coefficients
calculated from Eq.(\ref{basic-eq}).
Here the ``reflection'' and ``transmission'' waves are defined for the
quantum mechanics, whose ``space'' dimension is identified with $t$ in
Eq.(\ref{basic-eq}).
See Ref.\cite{TowardPR} for more details. 
Then, even if the first particle production is not given by the
conventional form
$n_{k,1}=e^{-\pi \kappa_1^2}$, a simple correspondence 
$\left.e^{-\pi \kappa_j^2}\right|_{j\rightarrow 1} \leftrightarrow
n_{k,1}$ can be used for the later calculation.
According to the method of Ref.~\cite{EMM}, the occupation number just after the first
particle production can be calculated as
\begin{equation}
n_{k,1} \sim \frac{2 \pi^2}{9} \exp{\left(-2.47 k^{3/2} \sqrt{\frac{\Lambda}{g \dot{\phi}^2}}\right)}.
\end{equation}
Note that the standard result $n_k \propto \exp
\left(-\pi\frac{k^2}{k_*^2}\right)$ is {\bf wrong} in this case. 
Of course this result is not trivial. See the original paper~\cite{EMM}
for more details.

One has to replace $k \rightarrow k/a(t_1)$ if one wants to introduce the expansion of the Universe. 
Then we can find that in Eq.(\ref{result1}) and (\ref{result1-n}), a replacement
\begin{equation}
e^{-\pi \kappa_j^2} \rightarrow \frac{2 \pi^2}{9}e^{-K_j}
\label{speculation}
\end{equation}
will give a reasonable estimation.
Since 
$K_j \equiv 2.47 \left(k/a(t_j) \right)^{3/2} \sqrt{\Lambda/(g\Phi_j^2
m_{\phi}^2)}\sim K_1$  
under the relations $\Phi_j \propto j^{-1}$ and $a_j\propto j^{2/3}$, 
Eq.(\ref{eq:nkj}) can be applied.
Then, the total energy transfer to $\rho_\chi$ can be calculated as
\begin{eqnarray}
\rho_{\chi,j}
&\sim&\frac{g\Phi_j^2}{\Lambda}\frac{1}{9a(t_j)^3}\sum_{I=1}^j{}_j{\mathrm C}_I
r^{I-1}\int^{\infty}_0 dk k^2
e^{-K_1 I}\nonumber\\
&\sim& \frac{2}{27\cdot (2.47)^2}\frac{g^2\Phi_1^2}{j^2\Lambda^2}\rho_{\phi,j}
\sum_{I=1}^j\frac{{}_j{\mathrm C}_Ir^{I-1}}{I^2} \nonumber \\
&\sim& \frac{2}{27\cdot (2.47)^2}\frac{g^2\Phi_1^2}{j^2\Lambda^2}
\rho_{\phi,j}\cdot\frac{(1+r)^{j+2}}{j^2r^3},
\label{eq:rho_chi_mu0}
\end{eqnarray}
where $r\equiv 4\pi^2/9$.
In the last line, the summation has been calculated using the method
mentioned below Eq.(\ref{eq-first-rho}), which is reliable for large $j$
but could have small deviation when $j$ is not large.
The last line of Eq.(\ref{eq:rho_chi_mu0}) is useful when one 
wants to apply the result to a non-inflaton field that may have
$\Phi_1\ll \Lambda$.
We can see that $\rho_{\chi,j}\sim\rho_{\phi,j}$ is achieved when 
\begin{equation}
j\sim 3.25+2\log_{1+r} j +2 \log_{1+r} \left(\frac{j\Lambda}{g\Phi_1}\right).
\label{eq:mu0_number}
\end{equation}
If we take $g\Phi_1/\Lambda=1$, energy transfer becomes significant at
$j\sim 8$. 
Neglecting the expansion of the Universe, we obtain 
$j\sim 5$, which is in good agreement with
our numerical calculation in the Minkowski space.

Looking into more details, we found that the numerical calculation
is still indicating that the growth factor of the number
density is seemingly larger than that of the analytical calculation. 
In our numerical calculation we found that the number density increases
as $9 \times 10^{-5} \rightarrow 5 \times 10^{-4} 
\rightarrow 2\times 10^{-3} \rightarrow 9\times 10^{-3} \rightarrow 3
\times 10^{-2} \rightarrow \cdots$.
The discrepancy could be caused by $\theta_j$ in Eq.(\ref{result1}) or
by Eq. (\ref{speculation}).
One will see similar excess in Fig.\ref{fig:comparing}. 
However, all these results are suggesting that preheating is efficient 
for non-renormalizable interaction, even though the oscillation is
``decoupled'' in the effective action.
Therefore, if chaotic inflationary model suggest $\Phi_1\sim
\Lambda$, preheating after inflation could be quite significant even if
the inflaton sector is ``decoupled''.
The above results do not depend  explicitly on the mass of
$\phi$ as long as $m_{\phi} \ll \Phi_1$.  
In contrast to the standard scenario, higher dimensional interaction
predicts significant dependence on the amplitude $\Phi$.

Our result can be applied to various other cases in
which a light scalar field (such as a curvaton or a modulus field)
begins oscillation after inflation.
It is also possible to identify $\chi$ as modulus field, and consider resonant
production after inflation. 
This case will be discussed in the next section.
Since particle production occurs in the area very close to the ESP, it
is possible to consider the case with $\Phi/\Lambda>1$.
Of course the effective action is not reliable when $\phi(t)/\Lambda>1$,
 but $\Phi/\Lambda<1$ is not always required for the calculation
as far as the particle production within the small area
$\phi(t)<\phi_*\ll \Lambda$ can be described using the effective
action.
In that way, a milder condition could be $m_\phi^2\Phi^2\ll\Lambda^4$.

\section{Intuitive argument for the decay rate}
The source of higher dimensional interaction could be diverse.
In Ref.\cite{EMM}, effective action of supersymmetric GUT model has been
discussed in detail.
If one focuses on Planck-scale suppressed interaction, it will be
helpful to consider terms like $\sim H^2\chi^2$.

In this section we consider ``decay rate'' for quadratic and quartic
potential to show how terms like $\sim H^2 \chi^2$ works in preheating.
To discuss the effective ``decay rate'', we are carefully following the
discussion of Sec.III and IV in Ref.\cite{TowardPR}. 
Since perturbative decay rate will appear in the $q\ll 1$ limit, 
it is useful to think about perturbative theory versus narrow resonance.

For the quadratic inflaton potential $V(\phi)\sim m_\phi^2 \phi^2/2$, 
we find effective interaction $H^2\chi^2\sim (m_\phi ^2/6M_p^2)\phi^2\chi^2$.
Although the source of this interaction is higher dimensional, the
resultant preheating uses ``conventional'' interaction $\sim g^2\chi^2\phi^2$
with $g^2\sim m_\phi^2/M_p^2\ll 1$. 
According to the argument in Ref.\cite{TowardPR}, perturbative
decay rate ($\Gamma_\phi$) could be significant when $\Gamma_\phi
>qm_\phi$, where $q\simeq 
\frac{g^2\Phi^2}{m_\phi^2}\sim \frac{\Phi^2}{M_p^2}$.
For a ``decoupled'' inflaton, considering the above effective
interaction $H^2\chi^2\sim (m_\phi 
^2/6M_p^2)\phi^2\chi^2$, $\Gamma_\phi$ can be estimated
as~\cite{TowardPR} $\Gamma_\phi(\phi\phi\rightarrow\chi\chi)
\sim g^4 \frac{\Phi^2}{m_\phi}\sim \frac{m_\phi^3\Phi^2}{M_p^4}$.
Also, it could be possible to consider light fermion interacting with 
${\cal L}_\mathrm{int}\sim(m_\phi/M_p)\bar{\psi}\psi \phi$.
Then one will find
$\Gamma_\phi(\phi\rightarrow\psi\psi) \sim \frac{m_\phi^3}{M_p^2}$.
These decay rates are suggesting that preheating is important as far as
$\Phi>m_\phi$. 

Next we consider quartic inflaton potential $V(\phi)\sim \lambda \phi^4/4$,
where $\lambda\ll 1$.
Again, one can expect higher dimensional effective interaction 
$c^2H^2\chi^2\sim c^2\lambda\phi^4\chi^2/(12M_p^2)$.
In this case preheating is
caused by truly higher dimensional interaction.
Since the oscillation occurs on quartic potential, the oscillating
solution is given by the elliptic function.
This case has been considered in Ref.\cite{quartic-pr} for
renormalizable interaction.
Repeating the argument of \cite{quartic-pr}, the elliptic function can be
replaced by the sinusoidal function.
Here we use the definitions used in Ref.\cite{quartic-pr}.
The mode equation for $X_k(t)=a(t)\chi_k(t)$ with the dimensionless
conformal time $x\equiv (48\lambda M_p^2)^{1/4}t^{1/2}$ (rescaled
using the amplitude $\Phi$) is
\begin{eqnarray}
X_k''+\left[\kappa^2
       +\frac{c^2\Phi^2}{12M_p^2}f(x)^4\right]X_k&=&0,
\end{eqnarray}
where $f(x)=cn\left(x,\frac{1}{\sqrt{2}}\right)$.
Here we are using equations (13) and (18) of Ref.\cite{quartic-pr}.
Denoting the period of the oscillation by $T$, one can expand 
\begin{eqnarray}
f^4(x)&=&\left[F_0+F_1\cos\left(\frac{4\pi x}{T}\right)
 +F_2\cos\left(\frac{8\pi x}{T}\right)+...\right]^2\nonumber\\
&=& \left(F_0^2+\frac{F_1^2}{2}+...\right)
+ 2F_0 F_1\cos\left(\frac{4\pi x}{T}\right)\nonumber\\
&&+\left(\frac{F_1^2}{2}+2F_0F_2\right)\cos\left(\frac{8\pi x}{T}\right)
+...,
\end{eqnarray}
where $F_0\simeq 0.46$, $F_1\simeq 0.50$ and
$F_2\simeq 0.04$.
See also Eq.(42) of Ref.\cite{quartic-pr}.
In units of $x$, the effective frequency is $2\pi/T\simeq 0.8$.
Specific value of $T$ is discussed below Eq.(14) of Ref.\cite{quartic-pr}.
Note that we are considering the same oscillation of $\phi(t)$
as Ref.\cite{quartic-pr}.
Then, considering ``only'' the leading term one can recover the Mathieu
equation, in which $q$ can be estimated as 
$q\sim c^2\frac{F_0 F_1\Phi^2}{12M_p^2}\left(\frac{T}{2\pi}\right)^2$.
This estimation is of course not rigorous, but would be useful in finding
a sensible estimation. 
Again, as far as the above approximations are valid, perturbative decay
from higher dimensional interaction ($H^2\chi^2\sim \lambda\phi^4\chi^2/12M_p^2$)
is not significant compared with preheating.
It is not obvious how fermions interact with $\phi$, since the source of
tiny $\lambda$ in the inflationary model is not quite obvious.
Even though, it could be
possible to consider interaction given by 
${\cal L}_\mathrm{int}\sim \lambda^{1/2} \bar{\psi}\psi
\phi^2/M_p$ to find that perturbative decay is not significant.

Because of fine-tunings of parameters required for inflation (e.g, $m_\phi\ll
M_p$ or $\lambda \ll 1$ in the above cases), it is not
quite obvious in reality how inflaton interacts with other fields.
The strength of the specific interaction could depend on the mechanism or the
symmetry that makes those parameters fine-tuned.
On the other hand, the interaction considered above ($\sim H^2 \chi^2$) 
could be mandatory.
Although intuitively, the above argument is showing importance of higher
dimensional interaction in inflationary cosmology.

\section{Conclusion and discussion}
\label{conclusion}
In this paper we considered a model of preheating when the oscillation
is ``decoupled''.
Even if the oscillation is ``decoupled'' in the effective action, higher
dimensional interaction could not be avoidable.
Oscillation in such model is normally a decoupled oscillation,
which has not been expected to cause efficient preheating.
In contrast to the usual expectation, we found that preheating could be
quite efficient for higher dimensional interaction.
Using three different approaches (quadratic approximation, steepest
descent method~\cite{EMM} and numerical calculation), we confirmed that
higher dimensional interaction can cause resonant particle production.
Unlike standard scenario of renormalizable preheating, the result
depends on the ratio $\Phi/\Lambda$.
Note that $\Phi/\Lambda>1$ is not excluded as far as the particle production
near the ESP is well described by the effective action.
The energy transfer is quick if the amplitude of the oscillation is
$\Phi_1\sim M_p$.
Let us answer to the ``naive questions'' given in the first section.
\begin{enumerate}
\item The final result contains $\Phi^2/\Lambda^2$, but it is wrong to
      think that $\phi(t)/\Lambda\ll 1$ causes suppression around the ESP. 
\item Again, although the result contains $\Phi^2/\Lambda^2$, it is
      wrong to think that higher dimensional interaction is negligible
      in preheating.
\item The function of $n_k$ is different. 
      Rigorous calculation is required to determine numerical
      coefficients. Since there is no exact solution, results have to be
      backed by numerical calculation. 
\end{enumerate}

Our result may also indicate that moduli oscillation could cause resonant
particle production even if the moduli is decoupled from other
particles.
More detailed study including instant preheating (i.e, when $\chi$ decays
during oscillation) and the curvaton oscillation in thermal environment will be
discussed in forthcoming paper.

\section*{Acknowledgement}
S.E. is supported in part  by the Polish NCN grant DEC-2012/04/A/ST2/00099.
N.M. is supported in part by Grants-in-Aid for Scientific Research from MEXT of 
Japan.

\end{document}